# Engineering the light coupling between metalens and photonic crystal resonators for robust on-chip microsystems

**Yahui Xiao[a], Zi Wang[a], Feifan Wang[a], Hwaseob Lee[a], Thomas Kananen[a], Tingyi Gu[a,*]**
[a] University of Delaware, Electrical and Computer Engineering, Newark, DE, USA, 19711

**Abstract**. We designed an on-chip transformative optic system of a metalens-photonic crystal resonator metasystem on a foundry compatible silicon photonic platform. By adjusting the on-chip metalens' focusing length and mode dimension, the insertion loss between the metalens and the photonic crystal resonator and waveguide structures are minimized through mode-matching. The micro-system does not involve any single mode silicon nanowire waveguide, and thus mechanically robust without any oxide claddings. The proposed microsystem is ideal for miniaturized chemical and biosensors operating in air or solution environment.

*Tingyi Gu**, E-mail: tingyigu@udel.edu

## 1 Introduction

The highest quality factor ($Q$) versus modal volume ($V$) ratio makes the photonic crystal (PhC) cavity an ideal platform with maximized light-matter interactions towards nonlinear optics, quantum optics, and sensing applications[1-8]. In conventional PhC circuits, the light is fed through a low-loss channel waveguide (WG) and a PhC WG for coupling into a PhC cavity[9,10]. The implementation of such a system has a few challenges: (1) Channel WG is usually fragile and easy to break during the undercut process, which significantly reduces yield during the postprocessing procedures[11], and make such a system less practical for any sensing applications; (2) The interface geometry between channel WG and PhC WG needs to be carefully engineered to minimize the insertion loss[12-16]; (3) As a line defect in PhC triangular lattice, the WG geometry needs to be engineered to ensure the overlap between the high-transmission band and PhC resonance wavelength. Here we propose a channel WG-free on-chip micro-system composed of a broadband dielectric metalens and a PhC resonator structure. The critical dimension of such a geometry is more than 100 nm and compatible with foundry processing[17,18]. The microsystems' mechanical robustness and foundry compatibility promise their applications in ultrafast low power modulators, hybrid lasers, and miniaturized biosensors.

In this work, we compare two channel WG-free designs with and without PhC WG. The direct coupling between a broadband metalens and PhC L3 cavity can lead to a 0.68 on-resonance transmission and a quality factor of 6,100, compared to the metalens – PhC WG side-coupled L3 cavity structure with the transmission of 0.66 and quality factor of 6,600. The correspondent extinction ratios are 84dB and 39dB, respectively.

## 2   Metalens – PhC WG with Side-coupled L3 Cavity System

This section focuses on optimizing the design of metalens (e.g., focal length and spot size[19,20]) and the PhC WG interface geometry to maximize the coupling efficiency. The interface of the PhC structure is designed by removing etched air holes to generate a PhC coupler with 38°, 60°,120°, and 180° (without removing air holes), respectively. Via sweeping the distance between the metalens and the PhC WG, the optimized position of the PhC structure can be computed from the 2.5D variational FDTD simulations.

*2.1 Design Principle of Low Loss On-chip Metalens*

Both metalens and PhC are defined on the same device layer on the silicon-on-insulator (SOI) substrate. The metalens is used as a compact low loss mode convertor[21]. Here we use a gradient varying high contrast transmit array (HCTA) metalens for wavefront control. The 1D metalens along the y-direction imposes a space-dependent phase shift on the TE polarized impinging light along the x-direction. Here, TE polarization is defined as the main E filed component is in the $y$-direction shown in Fig 1(b). The input light centered at the wavelength of 1.55µm propagates along the +x direction. The designed metalens is 10µm wide in the $y$-direction, with a focusing length of 13μm and a spot size of 0.65μm shown in Fig. 1(a), and the optical intensity distribution of focusing point at the $x - y$ plane in the 250-nm-thick silicon slab center. The spot size is marked as the full-width half-maximum (FWHM) in the cross-section of the mode profile in Fig. 1(c). As the focusing length increases from 4.6 to 25μm, the focusing efficiency increases from 29% to 78%, shown in Fig. 1(d). The focusing efficiency is defined as the fraction of the input light that passes through a rectangular aperture at the focal plane, with its width equals three times the spot size and a height of 0.5μm. The dashed line in Fig. 1(d) is the FWHM of the electric field intensity distribution in the PhC WG as a standard line for mode matching between metalens and PhC structure. The light with mode profile matching propagates smoothly from the focused Gaussian beam into the periodic PhC structure[22-24].

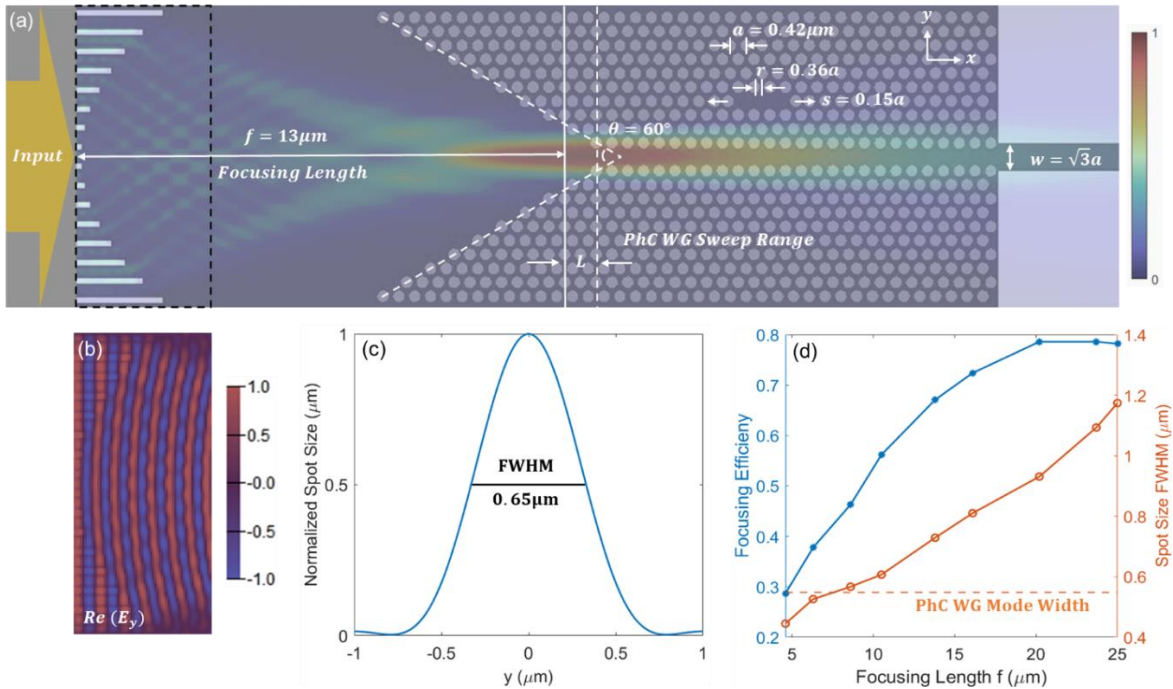

**Fig. 1** Mode matching between integrated metalens and PhC WG. (a) Schematic of the metalens – PhC WG system. (b) Simulated electric field Ey in the dashed box of (a). (c) The cross-sectional profile of light distribution on the focal plane, with the spot size marked as FWHM. (d) Focusing efficiency (blue line) and FWHM (orange line) of metalens on the focal plane versus focusing length, compared to the mode width of W1 PhC WG (orange dash line).

## 2.2 PhC WG Coupler

The structure of W1 PhC WG (relied on a single line-defect in hexagonal lattice) consists of air holes with a lattice constant a = 0.42μm etched into a silicon slab. The slab thickness is $d = 0.6a$ and air hole radius r = 0.36a, with the end holes shifted by s = 0.15a in a side-coupled L3 cavity shown in Fig. 1(a). The displaced end holes are primarily to increase the cavity volume for high-index-material (Si) to accumulate more photons, wherein the Bragg reflection and out-of-plane provide the confinement in-plane by total internal reflection (TIR). The width of the output channel WG is designed as w = $\sqrt{3}$a to ensure a high and stable transmission through the PhC WG. The channel WG position is swept to optimize the coupling efficiency between the PhC WG and channel WG. For a single-mode channel WG, the spatial mode profile is independent of the position along the WG direction. However, the spatial mode profile in PhC WG is changed by the periodic geometry along with the WG. In such a case, the coupling efficiency may change when the PhC WG ends with a different position within its period[25].

Via integrating with metalens (f = 13μm), the input light centered at the wavelength of 1.55μm is focused on spot size similar to PhC WG mode, which allows low insertion loss at the interface. Here, we designed four different PhC couplers with an angle of 38°, 60°, 120°, and 180° by removing individual air holes at the front edge of PhC WG to generate an input PhC taper[26-27]. Fig. 2(a) shows the optimized geometric structure and mode profile at the highest transmission region of PhC WG. The in-plane spot size of the given metalens is 0.65μm, which is smaller than the PhC coupler size (1.88μm) at the coupling plane.

Fig. 2(c) and (d) show the comparison of three different PhC coupler angles for total Q-factor and enhancement factor of side coupled L3 cavity. The total Q-factor of a resonator is solely limited by out-of-plane radiation that gives a direct measure of the cavity resonance lifetime, which can be re-expressed as $Q_{tot} = \frac{\lambda_0}{\Delta\lambda}$, where $\lambda_0$ is the resonant wavelength and $\Delta\lambda$ is the FWHM of the resonance. The enhancement factor implies the electric field intensity in the cavity at resonance.

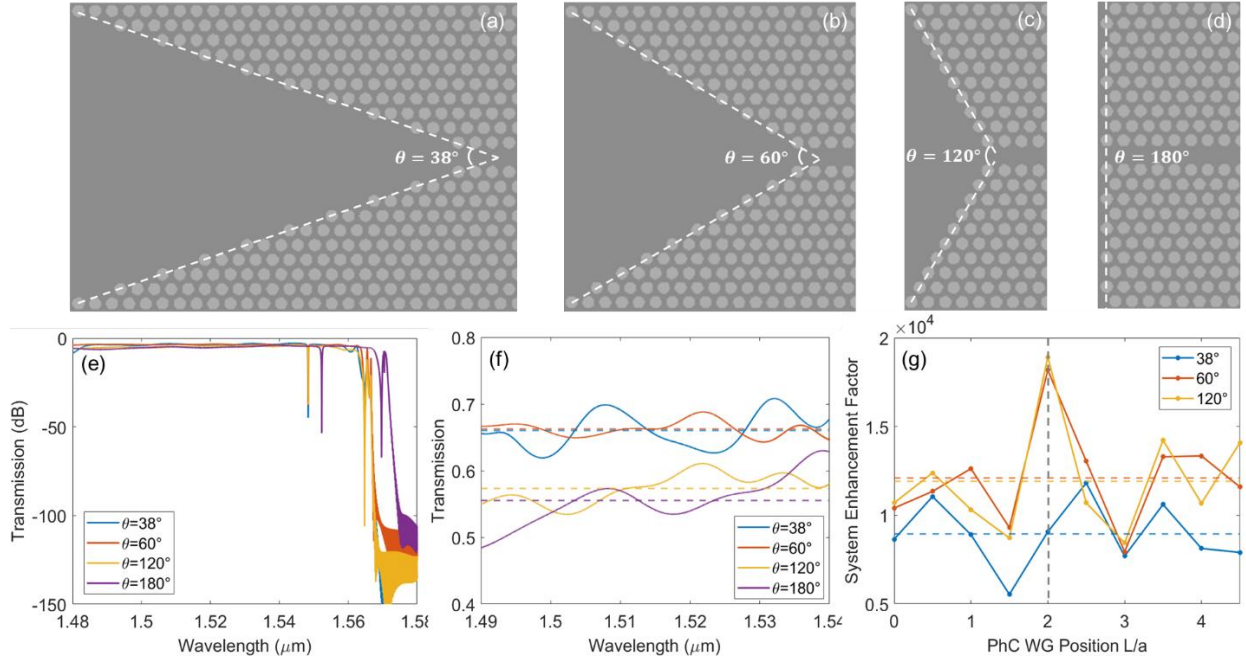

**Fig. 2** Mechanically robust metalens – PhC WG coupler. (a), (b), (c), and (d) are the geometries of PhC WG coupler with the angle of 38°, 60°, 120°, and 180°, respectively. (e) Transmission (Loss) with resonance dip of PhC WG of four different coupler angles in log scale. (f) Zoomed high transmission part of PhC WG of four different coupler angles in a linear scale. Dashed lines are the average transmission value for each coupler angle. (g) Sweeping system enhancement factor of PhC WG L3 cavity with coupler angle of 38°, 60° and 120°, with corresponding average values shown in dash lines.

By sweeping the distance between metalens and PhC WG (coupler angle of 60°) with a step size of 0.5a, the system enhancement factor shows an optimized distance between metalens and PhC WG of $f + L$, where $L = 2a$. However, PhC WG transmission is not quite sensitive to the changing distance based on sweeping results.

Fig. 2(e) shows the transmission of four different PhC WG coupler angles in log scale with resonance dips, where the side-coupled L3 cavities are excited by evanescent coupling from a nearby PhC WG[28]. The band edge of PhC WGs with the angle of 38°, 60°, and 120° is around 1.57μm. However, both resonance dip and the band edge of these three coupler angles shift to a slightly shorter wavelength, compared to the 180° coupler angle (without removing air holes). This phenomenon indicates that changing the PhC WG structure by removing air holes will shift the band edge to a shorter wavelength. Fig. 2(f) shows a zoomed transmission part from Fig. 2(e) in linear scale, where the coupler angle of 38° and 60° has the relatively highest transmission, while the one of 60° is slightly higher and more stable. Dashed lines are the average transmission value in the same color as the corresponding solid lines. The high transmission comparison shows that coupler with smaller angles can collect more light and decrease the back reflection by mode matching. Fig. 2(g) represents the PhC WG position sweeping results regarding system enhancement factor ($= \frac{I_{cavity}}{I_{input}}$), with the coupler angle of 38°, 60°, and 120°, respectively. Coupler angles of 60° and 120° show a similar trend and reach the optimized PhC WG position with $L = 2a$, marked by a vertical grey dashed line in Fig. 2(g).

## 3   Metalens – PhC L3 cavity System

Ln cavities represent line defects along $\Gamma - K$ direction, whereby n adjacent holes are removed from the periodic lattice to localize light along a line. The PhC L3 cavity is the most common configuration line-defect cavity (with three missing holes), with end holes shifted by 0.15a to obtain a high-Q factor[29], shown in Fig. 3(a) and (f). Thus, light is considered to penetrate more inside the mirror and be reflected perfectly, which means that the cavity edge's electric field profile becomes gentler. Here, we designed a PhC L3 cavity with the same parameters ($a = 0.42$μm, $r = 0.36a$, $s = 0.15a$) as the ones of PhC WG in Sec. 2. Similar to the metalens – PhC WG system, we also compared the PhC L3 cavity with the coupler angle of 38°, 60°, 120° by integrating with the same metalens ($f = 13$μm). The y-direction width of the PhC L3 cavity is designed as the same size of metalens to increase the coupling region and extend Bragg reflection in $x - y$ direction[30].

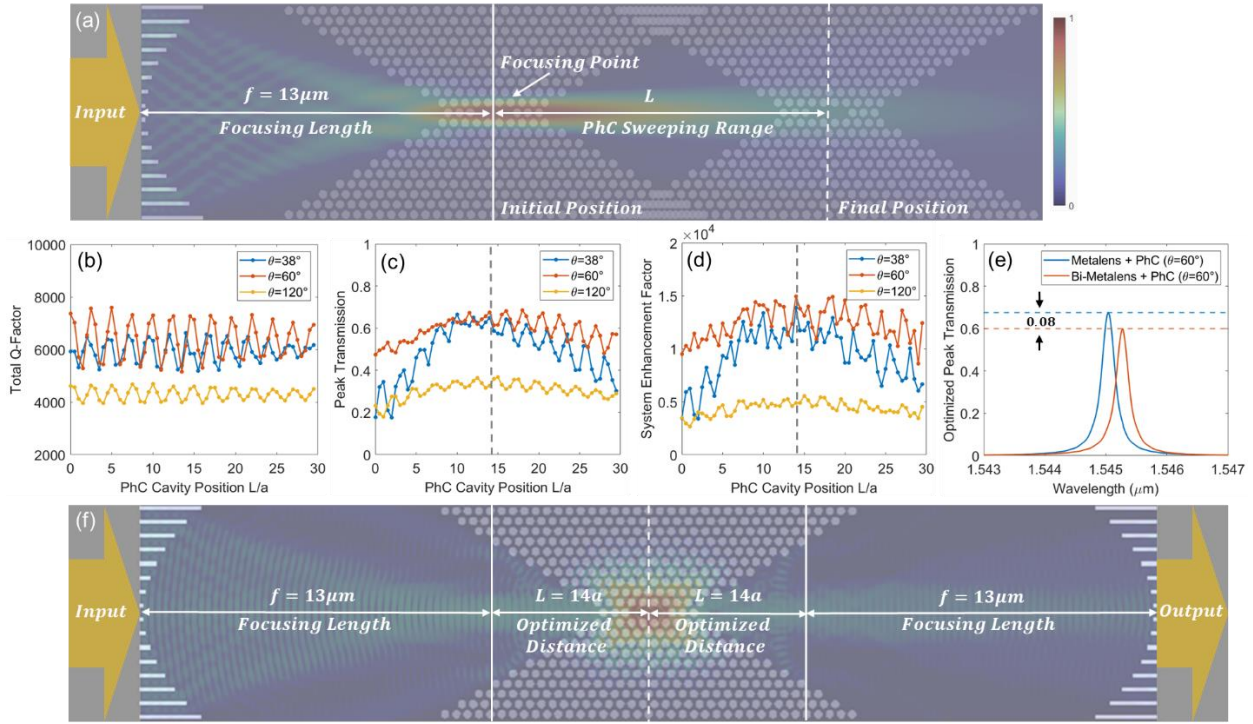

**Fig. 3** Mechanically robust metalens – PhC L3 Cavity coupler. (a) Device geometry of metalens and PhC L3 cavity. The distance between metalens and the center of the PhC L3 cavity is $f+L$. (b), (c), and (d) are the sweeping total Q factor, sweeping peak transmission, and the sweeping system enhancement factor of the device in (a). (e) Peak transmission comparison spectra of the device in (a) and (f) at an optimized distance, with PhC L3 cavity coupler angle of 60°. (f) Device geometry of metalens – PhC L3 cavity – metalens system with mode profile at resonance.

## 3.1 Sweeping Position of PhC L3 Cavity

In Fig. 3(a), the sweeping position of the PhC L3 cavity center starts from the focusing point (vertical dashed line) of the integrated metalens moving towards $+x$ direction, with a step size of 0.5a for a total of 60 points ending by the vertical solid line as the final position. The TE polarized light centered at 1.55μm converge through the metalens and impinge on the PhC L3 cavity to excite a fundamental resonant mode[31,32] at a specific wavelength by the tunneling wave. The simulated result for the total Q-factor in Fig. 3(b) shows stable but periodic oscillations, which confirms that during scanning Fabry-Pérot (FP) resonator length, the Airy distribution originates in the sum of mode profiles of the longitudinal resonator modes[33]. Based on the sweeping results in terms of the total Q-factor, the coupler angle of 60° shows the highest average value, slightly higher than the coupler angle of 38°. Fringes in Fig. 3(a) also implies the standing waves from interference in the FP resonator composed of metalens and the PhC structure.

In Fig. 3(c) and (d), the peak transmission and system enhancement factor appear an optimized PhC L3 position around L = 14a marked by the vertical grey dashed lines, a similar curvy oscillation trend at an optimized PhC cavity position range. The coupler angle of 60° (red data curves) shows the highest average data value in peak transmission and system enhancement comparison. The cavity resonance mode profile in the log scale shown in Fig. 3(f) may indicate that the mode direction of both left and right sides of the resonance mode contour appears 60°, which matches the PhC L3 cavity coupler angle of 60°.

*3.2 Metalens – PhC L3 Cavity – Metalens System*

Based on the sweeping results from Fig. 3(a), we selected the PhC cavity coupler angle of 60° at the optimized PhC cavity position (L = 14a) and added a second metalens (f = 13μm) symmetric about the center of the PhC L3 cavity based on the ray diagram of double lenses shown in Fig. 3(f). We also simulated the bilateral metalens system for the same PhC L3 cavity with the coupler angle of 38°, 60°, and 120°. As shown in Fig. 3(e), the highest peak transmission of the system is around 0.6 by the coupler angle of 60°, which is slightly lower than the one of metalens – PhC L3 cavity system around 0.68 in Fig. 3(a), mainly because of the transmission loss through the second metalens.

## 4   Systems Transmission Comparison

In this section, we compare the two complete integrating metalens system. Fig. 4 shows the transmissions of metalens – PhC WG system and the peak transmissions metalens – PhC L3 cavity – metalens system, with three coupler angles of 38°, 60°, 120°. For the transmission of metalens – PhC WG system, the coupler angle of 38° obtains the highest transmission ∼0.66 (marked in the blue curve), while for the bilateral metalens – PhC L3 cavity system, the highest peak transmission ∼ 0.6 achieved by 60° coupler angle (marked in the red dashed line). For the PhC structure with a coupler angle of 60°, the total Q-factor of the L3 cavity in the bilateral metalens system is around 6,000, while the total Q-factor in the PhC WG side-coupled L3 cavity is about 9,500. For the system enhancement factor, metalens – PhC WG obtains approximately 18,000, higher than the bilateral metalens – PhC L3 cavity with 13,000.

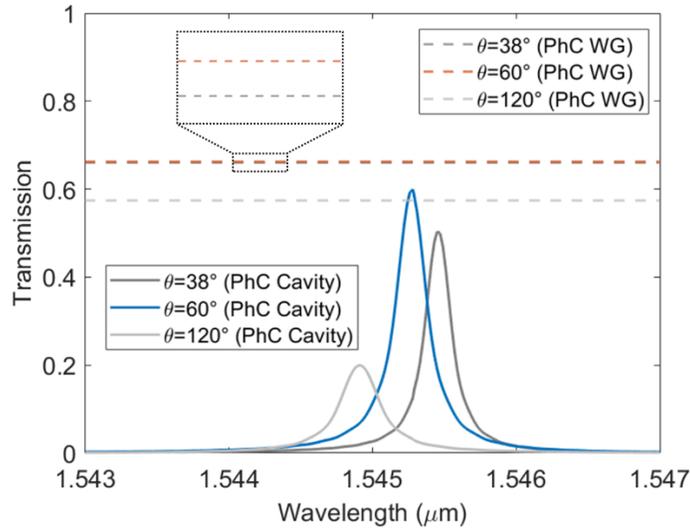

**Fig. 4** Average transmission spectra of the metalens – PhC WG system in Fig. 1(a) and peak transmission of metalens – PhC L3 cavity – metalens system in Fig. 3(f) with the PhC coupler angle of 38°, 60°, and 120°.

## 5 Conclusion

We demonstrate a low loss on-chip microsystem based on a broadband metalens and a PhC resonator through engineering the relative position and geometry of PhC and the metalens. Numerical examinations show that the coupler angle of 60° defined in the PhC WG interface leads to the lowest insertion loss (−4.2dB). While in the scheme of metalens – PhC L3 cavity, the coupler angle of 60° obtains the highest total Q-factor (∼6,100) and the system enhancement factor (∼15,000). The bilateral metalens system appears 0.08 lower than the metalens – PhC system concerning the peak transmission because of the loss from the second metalens.

The designed nanophotonic structures significantly reduce the footprint, reduce the insertion loss and improve the mechanical robustness of on-chip light coupling into PhC structure, which are desired for improving sensitivity in nanophotonic sensors[34-36] and reducing operation power for PhC based active silicon photonic components.

*Acknowledgments*

This work is supported by the AFOSR Young Investigator Program (FA9550-18-1-0300). Zi Wang is supported by the Early Career Faculty grant from NASA's Space Technology Research Grants Program (80NSSC17K0526).

**Yahui Xiao** is a Ph.D. candidate at the University of Delaware. She received her BS degree from the Changchun University of Science and Technology in 2018. Her current research interests include photonic crystals and metasurface. She is a student member of the Optical Society of America (OSA).

**Zi Wang** is a Ph.D. candidate at the University of Delaware. He received his BS and MS degree from the Beijing Institute of Technology in 2014 and 2017, respectively. His current research interests include a multi-layer meta-system and the application of the integrated photonic device for machine learning.

**Feifan Wang** has been a visiting student at the University of Delaware. She received her Ph.D. degree from Peking University in 2020. Her current research focuses on on-chip nanophotonic devices based on nonlinear effects.

**Hwaseob Lee** is a Ph.D. candidate at the University of Delaware. His research interests include several applications of nanophotonics based on silicon microring resonators. He is a student member of OSA.

**Thomas Kananen** obtained his MS and BS degree at the University of Delaware in 2018 and 2019, respectively. His research focuses on graphene plasmonic structures in a long wavelength. He is a student member of OSA.

**Tingyi Gu** is an assistant professor at the University of Delaware. She received her Ph.D. degree from Columbia University. Her research focuses on developing passive and active nanophotonic devices on a silicon photonic platform. She is a member of SPIE and OSA.